\documentclass[reprint, twocolumn, amssymb, amsmath, nobibnotes, aps, prresearch, superscriptaddress]{revtex4-1} 

\usepackage{graphicx}
\usepackage{float}
\usepackage{dcolumn}
\usepackage{bm}
\usepackage[dvipsnames]{xcolor}
\usepackage[colorlinks=true]{hyperref}
\usepackage{txfonts}
\definecolor{darkblue}{RGB}{33,33,137}
\definecolor{darkred}{RGB}{193,23,23}
\hypersetup{
    colorlinks=true,       
    linkcolor=blue,     
    citecolor=blue,    
    urlcolor=blue      
} 
\usepackage{natbib}
\usepackage[caption=false]{subfig}
\usepackage{color,soul}

\begin{document}


\title{Superconductor--Insulator Transition in a non-Fermi Liquid}

\author{A. L. Chudnovskiy}
\affiliation{
 1. Institut f\"ur Theoretische Physik, Universit\"at Hamburg,
 Notkestraße 9, D-22607 Hamburg, Germany
}%

\author{Alex Kamenev}

\affiliation{
 School of Physics and Astronomy, University of Minnesota, Minneapolis, Minnesota 55455, USA
}
\affiliation{
 William I. Fine Theoretical Physics Institute, University of Minnesota, Minneapolis, Minnesota 55455, USA
}%

\date{\today}

\begin{abstract} 
We present a model of a strongly correlated system with a non-Fermi liquid high temperature phase. Its ground state 
undergoes an insulator to superconductor quantum phase transition  (QPT) as a function of a pairing interaction strength. Both the insulator and the superconductor are originating from the same interaction mechanism. The resistivity in the insulating phase exhibits the activation behavior with the activation energy, which goes to zero at the  QPT. This leads to a wide quantum critical regime with an algebraic temperature dependence of the resistivity.  Upon raising the temperature in the superconducting phase, the model exhibits a finite temperature phase transition to a Bose metal phase, which separates the superconductor from the non-Fermi liquid metal. 
\end{abstract}

\maketitle


Quantum phase transition between the superconducting and insulating phases is observed in a variety of experimental systems, notably high-$T_c$ materials \cite{RevModPhys.84.1383,Met-Ins-SCdome2021,hidden-met-ins2021,PhysRevB.80.214531,Timusk_1999}, dirty superconducting films \cite{doi:10.1063/1.882069,PhysRevB.40.182,PhysRevLett.62.2180,RevModPhys.91.011002,doi:10.1073/pnas.1522435113,PhysRevB.93.100503}, and Josephson junction arrays \cite{PhysRevLett.63.326,PhysRevLett.63.1753,SC-INS_Exp2018}.  One concept  of the insulating phase is that of a system with a finite local superconducting order, but without the long-range phase coherence \cite{PhysRevB.22.459,PhysRevLett.45.1442,Efetov.JETP51.1015,PhysRevB.24.5063,Dubouchet2019,pseudogapLettNature2009}.  It presumes that Cooper pairs are dominant charge carriers for the electric transport not only in the superconducting but also in the insulating or metallic phases.  Remarkably, the high-temperature behavior of many of these 
systems is characterized as a strange metal with the linear in temperature resistivity.

\begin{figure}[htb]
  \centering
  \includegraphics[width=0.5\textwidth]{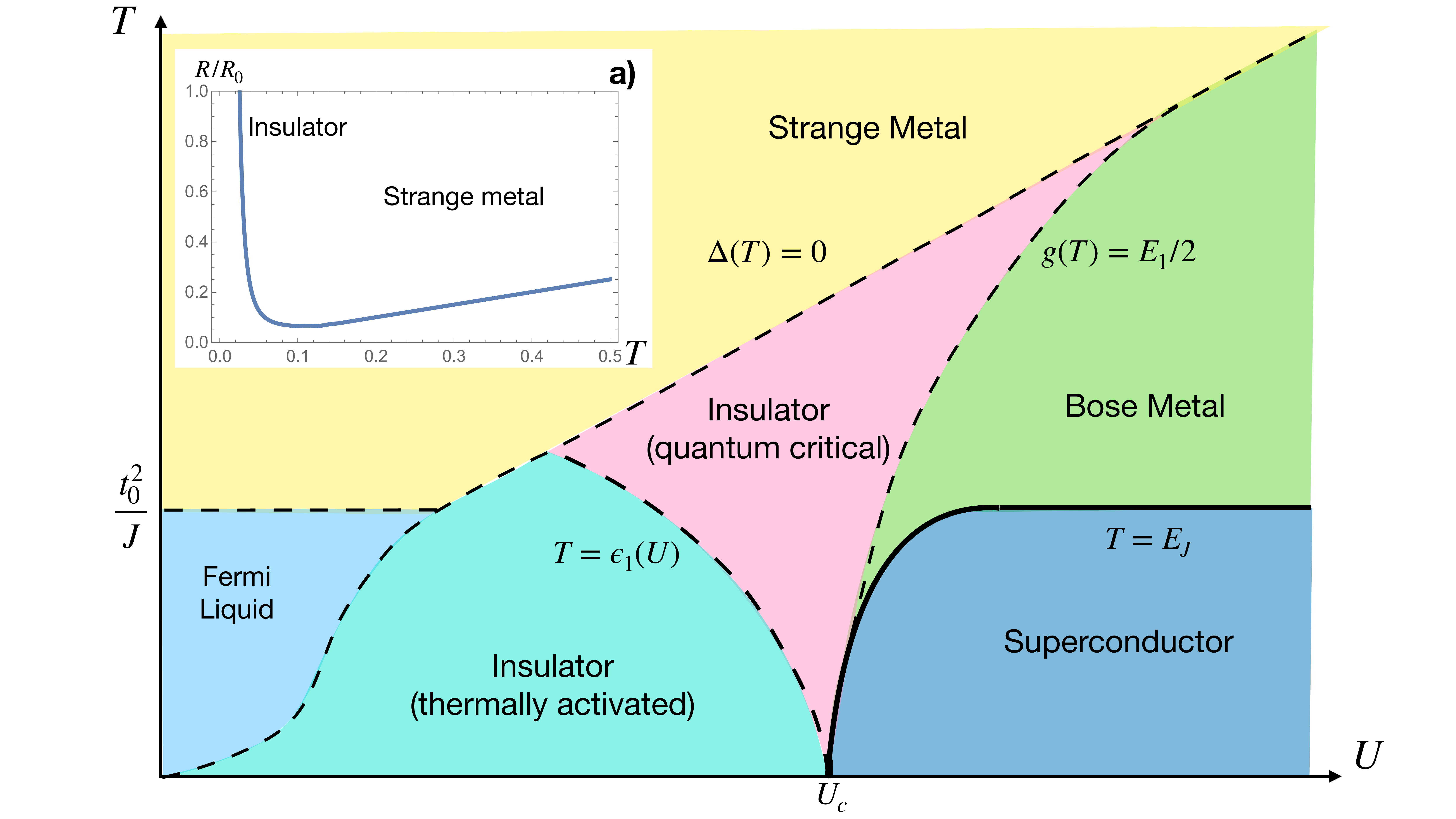}
  \caption{Sketch of the phase diagram of the SYK+U array. Inset a) Temperature dependence of the resistivity in the insulating (pseudogap) phase.}
  \label{fig:PhaseDiagram}
\end{figure}

Here we investigate a rather simple model, where the non-Fermi liquid strange metal state undergoes the superconducting instability upon lowering the temperature.
The origin of the superconductivity is in a certain pairing mechanism \cite{patel2018coherent,esterlis2019cooper,chowdhury2019intrinsic}, which we do not discuss in details and introduce phenomenologically as a negative Hubbard $U$ \cite{Wang2020}.  We show that the {\em same} pairing mechanism is responsible for the existence of  the insulating phase. The latter is separated from the superconductor by a quantum phase transition (QPT) at a critical value of the pairing strength, $U_c$, Fig.~\ref{fig:PhaseDiagram}. Within the insulating phase the low temperature resistivity exhibits the activation behavior, $R\propto\exp\{\epsilon_1/T\}$, with an activation energy $\epsilon_1(U)$.  Importantly, the activation energy tends to zero upon approaching the insulator to superconductor QPT, $\epsilon_1(U)\to 0$ when $U\to U_c-0$. This gives rise to a wide quantum critical regime above the QPT, which manifests itself as an insulator with the algebraically divergent resistivity. Yet the high-temperature resistivity in both the insulating and the superconducting phases exhibits linear behavior with the temperature $R\propto T$, attributed to the parent non-Fermi liquid state.

The model consists of an array of Sachdev-Ye-Kitaev (SYK) \cite{Sachdev1993,Kitaev2015}  grains, coupled through the single particle tunneling. 
Each grain contains $N\gg 1$ degenerate orbitals, with random four-fermion interactions.  
This model was thoroughly investigated in Refs. \cite{Song2017,PhysRevLett.123.066601,https://doi.org/10.48550/arxiv.2109.05037,https://doi.org/10.48550/arxiv.2203.04990,patel2018magnetotransport,chowdhury2018translationally},   where it was shown that its high temperature 
phase is a non-Fermi liquid metal with the linear in temperature resistivity.  To facilitate superconductivity we introduce a local attractive 
interactions between electrons with opposite spins.    The corresponding Hamiltonian is written as the sum of the 
intra-grain  SYK+U model and the inter-grain tunneling 
\begin{equation}
H=\sum_r H_{\mathrm{SYK+U}}^{(r)}+\sum_{\langle r, r' \rangle} H_{\mathrm{t}}^{(r,r')}. 
\label{SYK+U_array}
\end{equation} 
Here the first term describes the set of isolated  SYK+U grains labeled by $r$, 
\begin{equation} \label{eq:SYK-Grain}
  H_{\mathrm{SYK+U}}^{(r)}\! = \!\!\!
  \sum_{ijkl; \sigma\sigma^{\prime}}^N \!\!J_{ij;kl}^{(r)}  \,
  c_{r,i \sigma}^{\dagger} c_{r,j \sigma^{\prime}}^{\dagger} c_{r,k \sigma^{\prime}} c_{r,l \sigma}   
-  U\! \sum_{i}^N  c_{r,i \uparrow}^{\dagger} c_{r,i \downarrow}^{\dagger} c_{r,i \downarrow} c_{r,i \uparrow} ,  
\end{equation}
where $J^{(r)}_{ij;kl}$ is a real tensor with the following symmetry properties:
$ 
J_{ij;kl}^{(r)} = -J_{ji;kl}^{(r)} =-J_{ij;lk}^{(r)}=J_{lk;ji}^{(r)}   
$. 
The non-zero elements must have all four indexes $i,j,k,l$  distinct. Up to these symmetries, the matrix elements 
$J_{ij;kl}^{(r)}$ are assumed to be real independent random variables (uncorrelated between different grains), drawn 
from the Gaussian distribution with the zero mean, $\langle J_{ij;kl}^{(r)} \rangle=0$, 
and the variance    
$
\langle \big(J_{ij;kl}^{(r)}\big)^2 \rangle=J^2/(4N)^3
$. The last term in Eq.~(\ref{eq:SYK-Grain}) is the attractive Hubbard interaction, facilitating local on-orbital electron pairing.    

The spin conserving inter-grain tunneling is governed by the Hamiltonian 
\begin{equation}
H_{\mathrm{t}}^{(r,r')}= \sum_{ij;\sigma}^N t_{ij}^{(r,r')}\,  c^+_{ri\sigma}c_{r'j\sigma},
\label{Ht}
\end{equation}
where $t_{ij}^{(r,r')}$ are real Gaussian variables with zero mean and  $\langle \big(t_{ij}^{(r,r')}\big)^2\rangle=t_0^2/N$. Here $r, r'$ denotes nearest neighbor  positions of SYK+U grains within the array and $i,j=1,\ldots,N$ denote orbital labels within each grain.

Before tackling the array geometry, let us briefly remind the physics of a single SYK+U grain, as described by Eq.~(\ref{eq:SYK-Grain}),  \cite{Wang2020,patel2018coherent}. Its large-$N$ mean-field analysis results in the set of equations for the normal, $G$, and anomalous, $F$, Green functions, the corresponding self-energies, $\Sigma, \Xi$, and the self-consistent equation for the local (on-orbital) order parameter, $\Delta$:  
\begin{eqnarray}
&&   \hskip -.4cm
 G(\omega_n)=\frac{-i\omega_n+\Sigma(\omega_n)}{(\omega_n+i\Sigma(\omega_n))^2+(\bar{\Xi}(\omega_n)+\bar{\Delta})(\Xi(\omega_n)+\Delta)},  \label{SYK_G}
\\
 && 
\hskip -.4cm F(\omega_n)\!=\!\frac{-(\Delta+\Xi(\omega_n))}{(\omega_n+i\Sigma(\omega_n))^2+(\bar{\Xi}(\omega_n)+\bar{\Delta})(\Xi(\omega_n)+\Delta)}, 
\label{F_SP}\\
 &&  
\hskip -.4cm 
\Sigma_{\tau\tau'}=\frac{J^2}{32} G_{\tau\tau'}^3, \label{Sigma_SP}  
										 \, \,   \Xi_{\tau\tau'}=-\frac{J^2}{32}\bar{F}_{\tau\tau'}F_{\tau\tau'}^2,  \label{Xi_SP}   \\ 
&& 
\Delta=- U T \sum_{\omega_n} F(\omega_n)=- U F_{\tau\tau}. 
										\label{Delta_approx}
\end{eqnarray} 
Solution of Eqs.~(\ref{SYK_G}) - (\ref{Delta_approx}) shows that at $T=0$ there is  a finite local pairing  amplitude $\Delta(U)$  for any however small Hubbard $U>0$. In particular, in the weak-coupling BCS limit, $U\ll J$, it is given by $\Delta \sim J \exp\{-\sqrt{\pi}J/(8\sqrt{2}U)\}$. 
Solution of equations~(\ref{SYK_G}) - (\ref{Delta_approx}) also implies the hard energy gap, $\sim \Delta^2/J$, in the {\em single-particle} density of states.    

The mean-field treatment fails to account for fluctuations of the low-energy degrees of freedom represented by phases of local pairing amplitudes,  
$\langle c_{j \downarrow} c_{j \uparrow}\rangle= \Delta e^{i\phi_j}$,  with $j=1,\ldots,N$.  Dynamics of these phases is governed by an effective Hamiltonian 
 \begin{equation}
H_{K}=-E_1\sum_i^N \frac{\partial^2}{\partial\phi_i^2} -  \frac{g}{N}\sum_{i<j}^N \cos(\phi_i-\phi_j),
\label{HKuramoto}
\end{equation}
which represents the quantum version of the classical Kuramoto model of oscillator's synchronization \cite{Kuramoto1975,acebron2005kuramoto,dorfler2013synchronization,daido1992quasientrainment,
wiesenfeld1998frequency,witthaut2017classical,DSouza2019explosive,strogatz2000kuramoto,boccaletti2014structure,
gomez2007synchronizability,arenas2006synchronization}. Here $E_1\approx 0.52 J$ \cite{gu2019notes} and $g(U)\approx 16 \Delta^2(U)/J$ \cite{Wang2020}. 
The quantum Kuramoto model (\ref{HKuramoto}) exhibits a second order transition \cite{Wang2020} between the non-synchronized and synchronized ground states at the critical coupling $g_c=E_1/2$.  The corresponding global (within the grain) order parameter is given by  the ground state expectation value of the phase exponents $\left\langle \sum_j^N e^{i\phi_j}\right\rangle$. For $g>g_c$, it acquires a finite value, and the ground state of the Hamiltonian (\ref{HKuramoto}) possesses the off-diagonal long-range order.  Hence the grain is superconducting. For $g<g_c$, the orbital-specific superconducting phases are non-synchronized. Therefore, despite having a finite on-orbital order parameter $\Delta$, the ground state of the grain does not exhibit the long-range order and the superconductivity.  Given that the single-particle DoS is gapped, this leads to an insulating ground state.

This physics may be read off the correlation function of the order parameter 
$
\mathcal{D}_0(\tau-\tau')=\frac{1}{N}\sum_{i,j=1}^N\left\langle e^{i\phi_i(\tau)} e^{-i\phi_j(\tau')} \right\rangle
$. 
Its calculation in the non-synchronized (insulating) phase, detailed in the  Supplemental material \cite{Supplement}, results in 
\begin{equation}
\mathcal{D}_0(\omega_m)=
 \frac{2E_1}{\omega_m^2+\epsilon_1^2}, 
					 \label{expCorrelator}
\end{equation}
where 
\begin{equation}
\epsilon_1(U)=\sqrt{E_1\left[E_1-2g(U)\right]},  
							\label{epsilon1a}
\end{equation}
and $\omega_m=2\pi m T$ is bosonic Matsubara frequency. 
The correlation function Eq. (\ref{expCorrelator}) coincides with the Green function of a bosonic mode with the energy $\epsilon_1(U)$. The critical pairing strength $g_c=g(U_c)$ is found from the condition $\epsilon_1(U)=0$, as it indicates an instability of the non-synchronized phase towards a state with a non-zero global order parameter.  

The fact that the fluctuations of the order parameter exhibit a single resonant mode is a peculiarity of the Kuramoto model (\ref{HKuramoto}) with its all-to-all identical interactions. In more generic models there is a continuum of excitations. For example, substituting $g\to g_{ij}$ in Eq.~(\ref{HKuramoto}) and thinking of $g_{ij}$ as a connectivity matrix on some lattice or a graph, one finds a band  of Josephson plasmons \cite{PhysRevB.24.5063,PhysRevB.48.3316,PhysRevLett.56.2303}.   The important invariant feature, however, is that such a band is gapped from the ground state in the non-synchronized  phase, while the gap closing marks QPT to the globally ordered (superconducting) phase. 
 \begin{figure}[htb]
  \centering
  \includegraphics[width=0.5\textwidth]{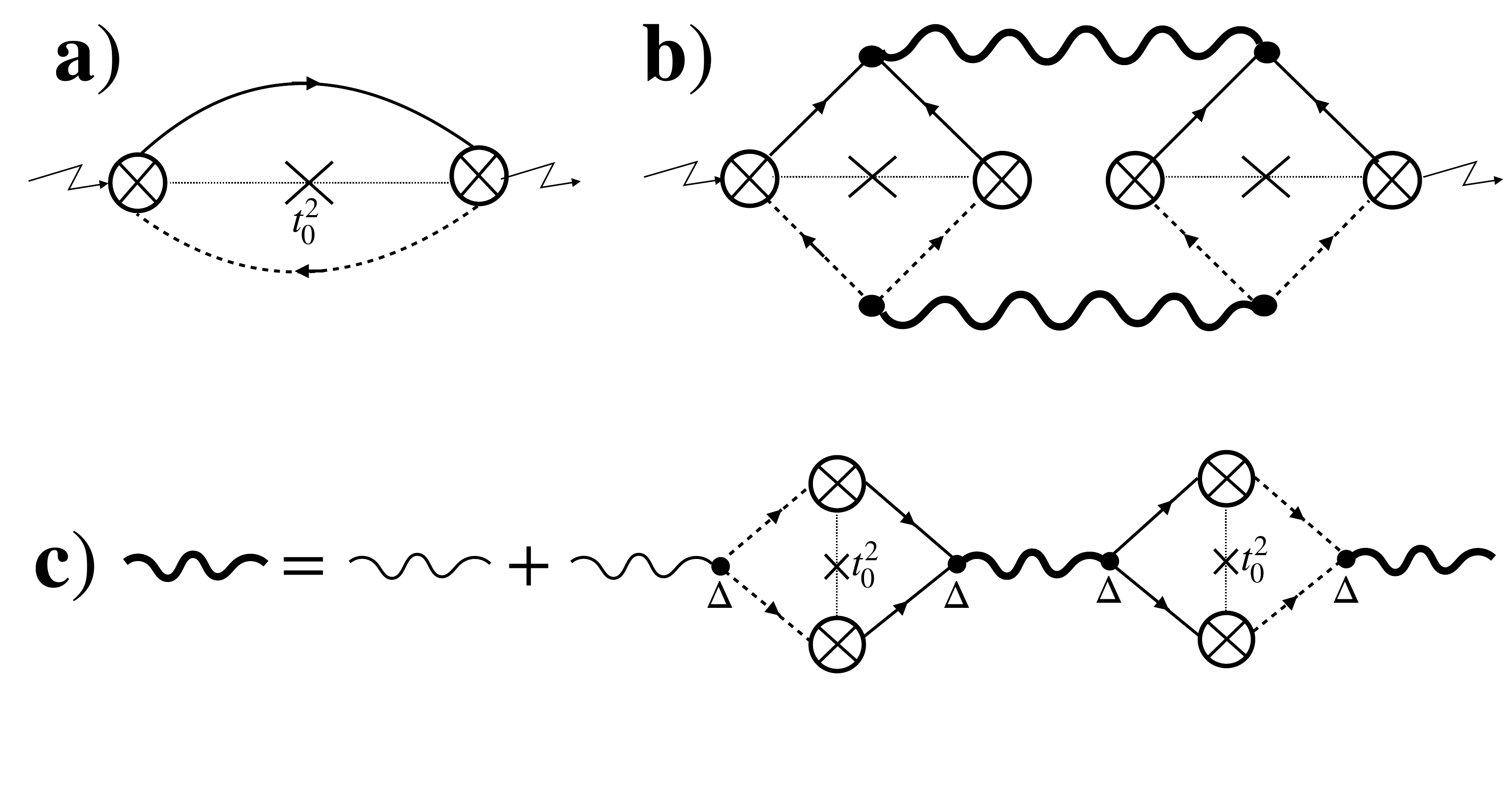}
  \vskip -.3cm
  \caption{a) One-particle normal conductivity; b) pair conductivity;    c) Dyson equation for the order parameter propagator. Thin and bold wavy lines denote the bare, Eq.~(\ref{expCorrelator}), and dressed in-grain order parameter propagator;  solid and dashed lines denote the one particle Green functions in different SYK-grains;  vertexes are given by tunneling amplitudes, and solid dots denote the superconducting amplitude. $\Delta$. Notice that the local self-energy, c),  involves tunneling of the pair out and back in the selected grain.}
  \label{fig:Amplitude}
\end{figure}

Having summarized the behavior of a single SYK+U grain, we turn now to our main subject -- the array of such grains. We start from the insulating 
phase, $U<U_c$.  Due to the Kuramoto phase desynchronization, the supercurrent between the grains is absent.     In the lowest (second) order in the tunneling amplitude, $t_{ij}^{(r,r')}$, the normal conductivity is given by the diagram 
Fig.~\ref{fig:Amplitude}(a). The corresponding normal Green functions in two neighboring grains are given by Eq.~(\ref{SYK_G}). Since the DoS is gapped with the gap $\Delta^2/J$, this contribution to the conductivity is exponentially suppressed with a factor $\exp\{-\Delta^2/JT\}$. 
Because the orbital pairing amplitude $\Delta$ is finite in the vicinity of the QPT, the single particle transport is strongly suppressed. 

This is not the case vis-a-vis transport of the pairs. The latter appears in the fourth order in the tunneling amplitude and is  given by the diagram 
Fig.~\ref{fig:Amplitude}(b), which is close in spirit to the Aslamazov-Larkin paraconductivity \cite{AslamasovLarkin68, ASLAMASOV1968238,Larkin-Varlamov,PhysRevB.97.014506}. The bold wavy lines there are propagators of the order parameter. Its bare form, given by Eq.~(\ref{expCorrelator}), is dressed by a bosonic self-energy, 
as shown in Fig.~\ref{fig:Amplitude}(c). Its imaginary  part, $\gamma$, accounts for a finite lifetime of a pair inside a grain due to tunneling to 
neighboring grains (the real part leads to a shift of the resonant energy $\epsilon_1$, which is of lesser interest).    Within our model $\gamma$ is 
found from the self-consistent solution  as $\gamma\sim\sqrt{ Z}\, E_J E_1/\epsilon_1$ for $\epsilon_1>Z^{1/4}\sqrt{E_J E_1}$, and $\gamma\sim  Z^{1/4}\sqrt{E_J E_1}$ for  $\epsilon_1 < Z^{1/4}\sqrt{E_J E_1}$ (for details see  Supplemental Material \cite{Supplement}) .  Here $Z$ is the coordination 
number of the array and the pair tunneling amplitude, $E_J \sim t_0^2/J$, given by the box in Figs.~\ref{fig:Amplitude}(b,c), represents the Josephson energy of the array (see Supplemental material for detailed calculations \cite{Supplement}). We note in passing, that $\gamma$ may include other mechanisms of the bosonic mode broadening, such as   
circuit noise or phonons. The intermediate formulas below account for those effects. 

The dressed retarded(advanced) propagators, are given by $\mathcal{D}^{R(A)}(\omega) =   \mathcal{D}_0(i\omega\mp \gamma)$.   Calculation of the diagram (b) in Fig.~\ref{fig:Amplitude} \cite{Supplement} results in the pair conductivity of the form   
\begin{equation}
\sigma=-\frac{(2e)^2}{h} E_J^2 \int_{-\infty}^{\infty}
\frac{ d\omega}{8T \sinh^2\left(\frac{\omega}{2T}\right) } \left(\mathcal{D}^R(\omega) -\mathcal{D}^A(\omega)\right)^2.
							\label{sigmaAL1}
\end{equation}
In the limit $T,\gamma  < \epsilon_1$,  one finds with the help of Eq.~(\ref{expCorrelator}) 
\begin{equation}
							\label{sigma_AL}  
\sigma =  \frac{(2e)^2}{h} \frac{8\pi E_J^2 E_1^2}{\epsilon_1^2 T\,  \gamma} \, e^{-\epsilon_1/T} \sim \frac{(2e)^2}{h} \frac{t_0^2}{\epsilon_1 T} \, e^{-\epsilon_1/T}.  
\end{equation}
Therefore the pair transport in the insulating state of the array exhibits the activation dependence on the temperature. 
The corresponding activation exponent, $\epsilon_1(U)$, Eq.~(\ref{epsilon1a}), is given by the gap in the fluctuation spectrum of the global order parameter. This gap goes to zero at the insulator to superconductor QPT, and therefore so does the activation energy of the pair conductivity.
This behavior should be contrasted with the single particle conductivity, Fig.~\ref{fig:Amplitude}(a).  The latter is also  activational, but its activation energy remains finite across the QPT.  As a result, the pair tunneling is the dominant transport mechanism close to the QPT. (Notice that, if the pair tunneling is the dominant mechanism of the mode broadening, $\gamma$, both the single-particle and the pair conductivity are proportional to $t_0^2$.)   

Equation (\ref{sigma_AL}) is not applicable in the quantum critical regime where $T>\epsilon_1$. In this case Eq.~(\ref{sigmaAL1}) yields conductivity which is a power-law in temperature 
\begin{equation}
\sigma\propto \frac{e^2}{h}  \frac{E_J^2 E_1^2 T}{\gamma^6} \times  \left\{ 
\begin{array}{ll}
T&\, \, \, \mbox{for} \, \,    T< \gamma, \\ 
\gamma  &\, \, \, \mbox{for} \, \,  T>\gamma.
\end{array}\right.
\label{sigmaAL_QCrit}
\end{equation}
These  power laws represent Gaussian exponents of the QPT. We do not attempt here to discuss if fluctuation corrections affect the critical exponents. The quantum critical regime extends above the $T=0$ QPT point,  Fig.~\ref{fig:PhaseDiagram}, flanked by the lines $T=\epsilon_1(U,T)$ on the left and $\epsilon_1(U,T)=0$ on the right. The insulating phase crosses over into the SYK strange metal phase with the linear in $T$ resistivity, at a temperature where the mean-field local pairing amplitude disappears $\Delta(T)=0$. This temperature marks closing the gap in the single-particle spectrum due to the 
meltdown of the preformed Cooper pairs.

The insulating phase with a  gap in the single-particle spectrum along with the preformed  pairs and strong phase fluctuations between the pairs on different orbitals may be considered as a model of the pseudogap state. The SYK physics stems from the fact that the interaction energy between these orbitals is larger than their kinetic energy dispersion. In the array geometry it naturally leads to the linear in temperature resistivity at high enough temperature \cite{Song2017}. This is a particular case of the recently developed theory of the Plankian metal, which emerges as the high-temperature phase of a conducting system without coherent quasiparticles   \cite{Song2017,PhysRevLett.123.066601,patel2018magnetotransport,chowdhury2018translationally,https://doi.org/10.48550/arxiv.2203.04990,https://doi.org/10.48550/arxiv.2109.05037,PhysRevB.100.045140,PhysRevB.103.195111}. At the same time, the strange metal theory is infra-red incomplete.  That is, there are many possible scenarios of the low-temperature phases that have a strange metal as their high-temperature limit \cite{RevModPhys.79.1015,doi:10.1146/annurev-conmatphys-031016-025531,JiangNFL2013,RevModPhys.92.031001}. 
The local orbital pairing phenomenology leads to the quantum Kuramoto model as its low-energy effective theory.  The latter  yields  
the insulating pseudogap phase with the activation resistivity and activation energy vanishing at the insulator to superconductor QPT.

We turn now to the superconducting side of the QPC. Here the ground state of each isolated grain is superconducting. The grains thus form 
the Josephson junction array with the inter-grain Josephson energy $E_J\sim t_0^2/J$. Since the individual phases of  $N$ orbitals within each 
grain are locked  $\phi^{(r)}_i=\phi^{(r)}$, the grain's charging energy is $E_C=E_1/N \ll E_J$. In other words, the effective electric capacitance of the
whole grain is $N$ times larger than  the quantum capacitance of the single orbital.     

By raising the temperature, the array undergoes the classical phase transition to the incoherent state at the critical temperature of the order of the Josephson energy $E_J$. At temperatures higher than the Josephson energy but still below the superconducting energy gap, the phase coherence between the grains in the array is lost, while each grain in the array  remains superconducting.  Therefore the single particle transport is suppressed and the normal current is
carried by incoherent Cooper pairs. It is thus denoted as Bose metal in Fig.~\ref{fig:PhaseDiagram}.

In Bose metal regime the conductivity of the array can be calculated from the conductivity of a single junction with simple circuit theory rules.  
We thus evaluate the conductivity of one junction, considering the rest of the array as an effective dissipative medium.  Within this approach an escape of a Cooper pair from a grain through $(Z-1)$ other junctions provides a mechanism for the loss of the phase coherence across the junction of interest.  The single Josephson junction is governed  by the action 
\begin{equation}
S_{JJ}=\int dt \left\{\frac{\dot{\phi}^2}{2E_C} +E_J \cos(\phi)\right\}. 
\label{S1J}
\end{equation}
The resulting equation of motion $\ddot{\phi}=E_C E_J\sin(\phi)$ describes  dynamics of the superconducting phase of an isolated junction. For a junction embedded in the incoherent  array, the equation of motion above is amended with the  relaxation term and the corresponding Langevin noise  
\begin{equation}
\frac{(2e)^2}{E_C}\ddot{\phi} + \sigma \dot{\phi}=(2e)^2 E_J \sin(\phi)+\xi(t). 
\label{LangevinPhase}
\end{equation}   
The dissipation coefficient $\sigma$ and the Langevin noise $\xi$ are tied  by the fluctuation-dissipation  relation
\begin{equation}
\langle \xi(t) \xi(t')\rangle=2 \sigma T \delta(t-t'). 
\label{FDT}
\end{equation}
Consequently $\sigma$ is identified with the Ohmic conductivity of the junction, which is in turn related to the circuit RC relaxation rate $\gamma$ as $\sigma=(2e)^2 \gamma/E_C$. The RC relaxation rate is given by a pair escape into the rest of the array 
evaluated  in the Supplemental material \cite{Supplement} as $\gamma=\sqrt{Z}E_J$. With this we obtain for the conductivity of the Bose metal 
\begin{equation}
\sigma\propto E_J/E_C \sim N\,\frac{t_0^2}{J^2}. 
\end{equation}
Parametrically it is the same as conductivity in the low temperature, $T< t_0^2/J$, Fermi liquid regime of the normal SYK array \cite{Song2017}. The physics, however, is very different and the Bose metal appears at higher temperature $T\gtrsim E_J\approx t_0^2/J$ and extends up to the in-grain critical temperature. The latter is determined  
by the condition $g(T)=16 \Delta^2(U,T)/J=E_1/2$, where the Kuramoto synchronization within the grain is lost. 
At higher temperature the array enters the quantum critical region described above, see Eq. (\ref{sigmaAL_QCrit}) and Fig. \ref{fig:PhaseDiagram}.  



In conclusion, we proposed an array of SYK+U grains as a microscopic model that reproduces some crucial parts of the high-$T_c$ phase diagram.  Despite simplicity of its Hamiltonian, the model exhibits a rather rich phenomenology. At 
zero temperature it features QPT between the insulating and the superconducting states, both induced by the Hubbard U. 
The insulating phase reminds the pseudogap physics of preformed incoherent Cooper pairs, described by the quantum Kuramoto picture. The low-temperature resistivity here obeys the Arrhenius law with the activation energy, which smoothly goes to zero approaching the transition.   At yet a higher temperature the Cooper pairs melt giving way to a non Fermi liquid metal with the linear in temperature resistivity.    This features qualitatively agree with a  non-monotonous temperature dependence of resistance in the pseudogap phase observed in several experiments \cite{annurev-conmatphys-070909-104117,PhysRevB.80.214531,PhysRevB.79.180505}.     On the superconducting side of the transition, upon elevating temperature, the system goes into the Bose metal and eventually again  to the non Fermi liquid metal. The latter phenomenology is relevant for Josephson arrays  \cite{SC-INS_Exp2018} and  
 for other systems featuring superconductor -- insulator transition, such as disordered superconducting films, where the transport of Cooper pairs provides the dominant charge transport channel \cite{doi:10.1126/science.aax5798,PhysRevB.40.182,RevModPhys.91.011002,TIKHONOV2020168138}.     



We are grateful to A. Chubukov,  L. Glazman, A. Goldman, C. Marcus,  and  A. Schnirman  for useful discussions. 
AK was supported by the NSF grant DMR-2037654.


\bibliography{SYK_superconductor.bib}

\onecolumngrid

\newpage

\section*{Supplemental Material}

\subsection{Generating functional for exponential correlation functions}
\label{sec:exp_phi}
Here we provide detailed derivation of the phase correlation functions Eq. (8) of the main text. 
The imaginary time action for the Hamiltonian Eq. (7) reads 
 \begin{equation}
 S[\phi_i(\tau)]=\int_0^{\beta}d\tau \left[\frac{1}{4E_1}\sum_{i=1}^N \dot{\phi}_i^2-\frac{g}{N}\sum_{i<j} \cos(\phi_i-\phi_j)\right].
 \end{equation}
This is the action for $N$ particles on the ring interacting by the cosine potential. We employ Hubbard-Stratonovich decoupling, introducing the order parameter field for the phase synchronization transition 
\begin{equation}
\rho=\frac{1}{N}\sum_{i=1}^N\left\langle e^{i\phi_i}\right\rangle, 
\label{Def_rho}
\end{equation}   
which results in the following partition function 
\begin{equation}
Z=\int[D \bar{\rho}^{\tau}, \rho^{\tau}]\exp\left[-N S[\bar{\rho}^{\tau}, \rho^{\tau}]\right], 
\end{equation}
where 
\begin{equation}
S[\bar{\rho}^{\tau}, \rho^{\tau}]=\frac{1}{g}\int d\tau \bar{\rho}^{\tau}\rho^{\tau}-\ln Z_{\phi}[\bar{\rho}^{\tau}, \rho^{\tau}], 
\label{action_rho}
\end{equation}
\begin{equation}
Z_{\phi}[\bar{\rho}^{\tau}, \rho^{\tau}]=\int D[\phi^{\tau}]\exp\left[-\int d\tau \left\{\frac{\dot\phi^2}{4E_1}-\left(\bar{\rho}^{\tau}e^{i\phi^{\tau}}+\rho^{\tau}e^{-i\phi^{\tau}}\right)\right\}\right]
\label{Zphi}
\end{equation}

To obtain the correlation function of the phase exponents at different sites, we introduce site-local source fields  
$\bar{\eta}_i^{\tau}, \eta_i^{\tau}$, thus obtaining the generating functional  
\begin{equation}
Z_{\eta}=\int[D \bar{\rho}^{\tau}, \rho^{\tau}]\exp\left[-\frac{N}{g}\int d\tau \bar{\rho}^{\tau}\rho^{\tau}   + \sum_{i=1}^N \ln Z_{\phi}[\rho, \eta_i]\right], 
\end{equation}
where 
\begin{equation}
Z_{\phi}[\rho, \eta_i]=\int[D\phi]\exp\left[-\int d\tau \left\{ \frac{\dot{\phi}^2}{4E_1}-(\bar{\rho}^{\tau}+\bar{\eta}_i^{\tau})e^{i\phi^{\tau}} 
-(\rho^{\tau}+\eta_i^{\tau})e^{-i\phi^{\tau}} \right\} \right]. 
\end{equation}
Further we expand $\ln Z_{\phi}$ up to quadratic order in $\rho$ in the non-synchronized phase ($\langle e^{\pm i \phi}\rangle=0$) to get 
\begin{equation}
\ln Z_{\phi}[\rho, \eta_i]\approx \ln Z_0+ \int d\tau d\tau' \left\langle e^{i\phi^{\tau}} e^{-i\phi^{\tau'}} \right\rangle (\bar{\rho}^{\tau}+\bar{\eta}_i^{\tau})  (\rho^{\tau'}+\eta_i^{\tau'}), 
\end{equation}
where 
\begin{equation}
Z_{0}=\int[D\phi]\exp\left[-\int d\tau \frac{\dot{\phi}^2}{4E_1} \right]. 
\label{Z0}
\end{equation}
Note that $Z_0$ is the partition function of a free quantum rotor, hence it can be calculated exactly using the energy spectrum $E_{\ell}=E_1\ell^2$, where $\ell$ denotes $z$-projection of the angular momentum.

Going over to Matsubara frequency space, the total generating function can be written as 
\begin{eqnarray}
\nonumber && 
Z_{\eta}=\int[D \bar{\rho}, \rho]\exp\left[-\sum_{\omega_n} \left\{ \bar{\rho}(\omega_n) \left[\frac{N}{g}- N D_0(\omega_n)\right] \rho(\omega_n) -
\right. \right. \\ 
\nonumber && \left. \left. 
D_0(\omega_n) \sum_{i=1}^N\left[\bar{\rho}(\omega_n)\eta_i(\omega_n)+ \bar{\eta}_i(\omega_n)\rho(\omega_n)\right]- 
 D_0(\omega_n) \sum_{i=1}^N\bar{\eta}_i(\omega_n) \eta_i(\omega_n)\right\}   \right], \\ 
\end{eqnarray} 
where $D_0(\omega_n)$ denotes the on-site correlation function of phase exponents in the absence of Kuramoto interaction, 
\begin{equation}
D_0(\omega_n)=\left\langle T_{\tau}\left( e^{i\phi^{\tau}} e^{-i\phi^{\tau'}}\right) \right\rangle_{ \omega_n}=  \frac{1}{\mathcal{N}}\sum_{\ell=0}^{\infty} D_{\ell}(\omega) \left[e^{-\frac{E_1}{T}\ell^2}-e^{-\frac{E_1}{T}(\ell+1)^2}\right],  
\label{D0}
\end{equation}
where $T_{\tau}$ denotes the time-ordering in the imaginary time $\tau$, 
\begin{equation}
D_{\ell}(\omega)=\frac{2E_1(2\ell +1)}{\omega_n^2+E_1^2(2\ell +1)^2}, \, \, \, \, \mathcal{N}=\sum_{\ell=-\infty}^{\infty} e^{-\frac{E_1}{T}\ell^2}.
\label{Dl_0}
\end{equation}
The normalization factor $\mathcal{N}$ ensures the condition $\left\langle e^{i\phi^{\tau}} e^{-i\phi^{\tau}} \right\rangle=1$.

At low temperatures, $T\ll E_1$, $D_0(\omega_n)$ can be approximated by the term with $\ell=0$,  
\begin{equation}
D_0(\omega_n)\approx \frac{2E_1}{\omega_n^2+E_1^2}.
\label{D0_lowT} 
\end{equation}
This approximation is used in the main text of the paper. 

Integrating out the fields $\bar{\rho}$, $\rho$, we obtain  
\begin{equation}
Z_{\eta}=\exp\left[\sum_{\omega_n}\sum_{i,j=1}^N \bar{\eta}_i(\omega_n) \mathcal{D}_{ij}(\omega_n)\eta_j(\omega_n)\right], 
\end{equation}
where 
\begin{equation}
\mathcal{D}_{ij}(\omega_n)=D_0(\omega_n)\delta_{ij} +\frac{1}{N}\frac{D_0^2(\omega_n)}{\frac{1}{g}-D_0(\omega_n)}
\label{Dij}
\end{equation}
Explicitly
\begin{equation}
\mathcal{D}_{ij}(\omega_n) = \frac{2E_1}{\omega_n^2+E_1^2}\delta_{ij}  + \frac{4E_1^2 g/N}{(\omega_n^2+E_1^2)[\omega_n^2+E_1(E_1-2g)]}.
\label{Dij_explicit}
\end{equation} 
Eq. (\ref{Dij_explicit}) gives the correlation functions of phase exponents, according to 
\begin{equation}
\left\langle e^{i\phi_i^{\tau}} e^{-i\phi_j^{\tau'}} \right\rangle_{ \omega_n}=
\frac{\partial^2 \ln Z[\bar{\eta}, \eta]}{\partial\bar{\eta}_i(\omega_n) \partial\eta_j(\omega_n)}= \mathcal{D}_{ij}(\omega_n).
\end{equation}
The second term in Eqs. (\ref{Dij}), (\ref{Dij_explicit}) diverges for $\omega_n=0$ at the critical point $g=E_1/2$, although the critical fluctuations enter the correlation function on two different sites as a $1/N$ correction only.  However, the following correlation function retains the critical behavior in the limit $N\rightarrow \infty$
\begin{equation}
\mathcal{D}(i\omega_n)=\frac{1}{N}\sum_{i,j}\left\langle e^{i\phi_i^{\tau}} e^{-i\phi_j^{\tau'}} \right\rangle_{ \omega_n}= 
 \frac{2E_1}{\omega_n^2+E_1(E_1-2g)}.
 \label{D_iomega}
\end{equation}
This is exactly the correlation function of phase exponents in Eq. (9) of the main text of the paper with $\epsilon_1=\sqrt{E_1(E_1-2g)}$ given by Eq. (10) in the main text. 

\subsection{One particle Green function}

The one particle Green function of the SYK+U model has been determined analytically in the limiting cases of low and high temperature in Ref. \cite{Wang2020}. At high temperature, the Green function retains the form of the pure SYK model without the additional Hubbard interaction. At low temperature, the presence of Cooper pairs leads to the gap in the one particle excitation spectrum, which has been evaluated  in Ref. \cite{Wang2020} as 
\begin{equation}
\Delta_1\approx 4\sqrt{6\pi}\frac{\Delta^2}{J},  
\label{Delta1}
\end{equation}
where $\Delta$ denotes the absolute value of the local Cooper pair amplitude given by the solution of the mean field equation. In contrast to the conventional BCS superconductor,  the one particle energy gap is strongly reduced due to the non-Femi-liquid ground state of the underlying SYK model.  The presence of the gap leads to the exponential decay of the one particle Green function with the decay time $\tau_{\Delta}=1/\Delta_1$ 
\begin{equation}
G(\tau)\approx 
-\left(\frac{8}{\pi}\right)^{1/4}\frac{e^{-|\tau|/\tau_{\Delta} }}{\sqrt{J|\tau|}}\, \mathrm{sgn}(\tau). 
\end{equation}
For further calculations we adopt simplified form of the one-particle Green functions, replacing the decay on the time-scale $\tau_{\Delta}$ by a hard cutoff. With that approximation, the imaginary time one particle Green function at zero temperature reads 
\begin{equation}
G(\tau)=-\left(\frac{8}{\pi}\right)^{1/4}\frac{\mathrm{sign}(\tau)}{\sqrt{J|\tau|}} \theta(\tau_{\Delta}-|\tau|), 
\label{Gtau}
\end{equation}
where $\theta(\tau_{\Delta}-|\tau|)$ denotes the Heaviside step function.

\subsection{Inter-grain Josephson coupling}
\begin{figure}[htb]
  \centering
  \includegraphics[width=0.3\textwidth]{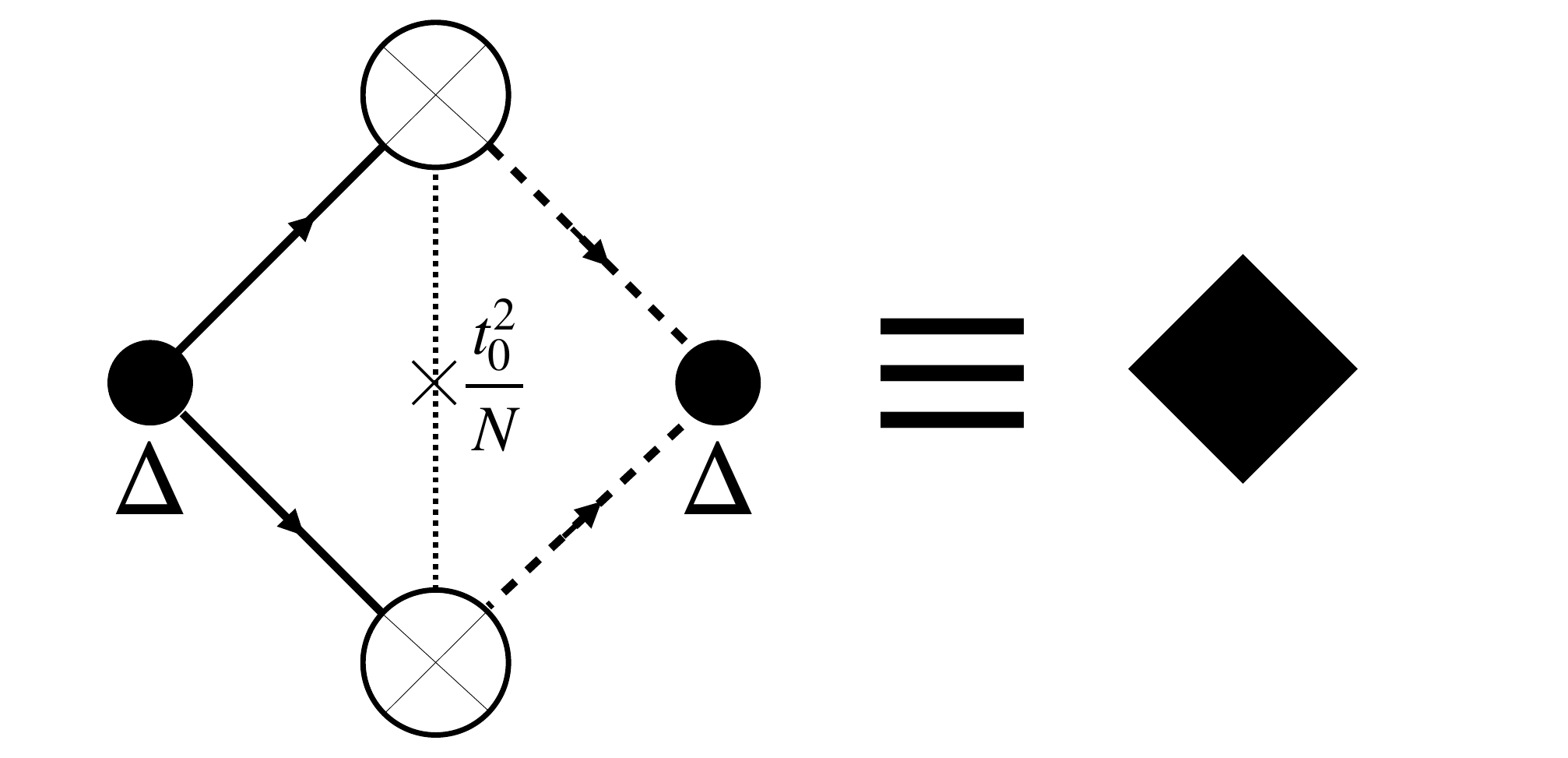}
  \caption{Vertex diagram for the Josephson coupling between the grains $r$ and $r'$. Solid lines denote one-particle Green functions in the grain $r$, dashed lines denote the one-particle Green functions in the grain $r'$. The circles denote the superconducting amplitudes $\Delta$.}
  \label{fig:JCoupling}
\end{figure}

In this subsection we present the approximate evaluation of the Josephson coupling between two SYK+U grains. The diagram for the interaction between the superconducting order parameters in the two grains is shown in Fig. \ref{fig:JCoupling}.  
The corresponding analytic expression in imaginary time action is given by 
\begin{eqnarray}
\nonumber && 
S_{\Delta\Delta}=\int d\tau d\tau'
\frac{t_0^2}{N}\sum_{i, j}^N\bar{\Delta}_{ri}(\tau) \Delta_{r'i}(\tau')\int d\tau_1d\tau_2 G(\tau-\tau_1)G(\tau_1-\tau')G(\tau-\tau_2)G(\tau_2-\tau')+c.c.\\ 
&& 
=\frac{1}{N}\sum_{i, j}^N\int d\tau d\tau' \bar{\Delta}_{ri}(\tau) \Delta_{r'j}(\tau') (\mathcal{P}(\tau, \tau'))^2 + c.c., 
\label{SDeltaDelta}
\end{eqnarray} 
where we defined 
\begin{equation}
\mathcal{P}(\tau, \tau')=t_0 \int d\tau_1 G(\tau-\tau_1)G(\tau_1-\tau').
\label{P}
\end{equation}
Evaluation of the integral in Eq. (\ref{P}) using the approximate Green functions Eq. (\ref{Gtau})  results in 
\begin{eqnarray}
\nonumber && 
\mathcal{P}(\tau, \tau')= \left(\frac{8}{\pi}\right)^{1/2}
\frac{t_0}{J} \left\{\pi 
-4 \mathrm{arsinh}\left[\sqrt{\frac{\tau_{\Delta}-|\tau-\tau'|}{|\tau-\tau'|}}\right] \theta(\tau_{\Delta}-|\tau-\tau'|)- \right. \\ 
&& \left. 
2\left[\arcsin\left(\sqrt{\frac{\tau_{\Delta}}{|\tau-\tau'|}}\right)
-\arctan\left(\sqrt{\frac{|\tau-\tau'|-\tau_{\Delta}}{\tau_{\Delta}}}\right)  
\right]\theta(|\tau-\tau'|-\tau_{\Delta})\theta(2\tau_{\Delta}-|\tau-\tau'|)
\right\}. 
\label{Pintegrated}
\end{eqnarray}
Taking into account that the integration kernel depends only logarithmically on the difference $\tau-\tau'$ with maximum at small time-differences, we can write the action Eq. (\ref{SDeltaDelta}) for $\Delta(\tau)$ varying at the time-scale much larger than   $\tau_{\Delta}$ in the approximately local form. Substituting Eq. (\ref{Pintegrated}) into Eq. (\ref{SDeltaDelta}), performing the integration over the time-difference $\tau-\tau'$,  and using the definition $\tau_{\Delta}=1/\Delta_1=\frac{J}{4\sqrt{6\pi}\Delta^2}$,  we obtain 
\begin{equation}
S_{\Delta\Delta}\approx 3.7  \frac{t_0^2}{N J \Delta^2}  \int d\tau \sum_{i, j}^N\bar{\Delta}_{ri}(\tau) \Delta_{r'j}(\tau)  + c.c. 
\label{SDeltaDeltafin}
\end{equation}
Finally, separating the superconducting order parameter at each site of each  SYK+U grain into the constant mean field amplitude and fluctuating phase, $\Delta_{ri}=
\Delta e^{i\phi_{ri}(\tau)}$, we can represent Eq. (\ref{SDeltaDeltafin}) in the form of the action for the Josephson junction 
\begin{equation}
S_{JJ}=\frac{E_J}{N}\int d\tau \sum_{i,j}^N \cos(\phi_{ri}^{\tau}-\phi_{r', j}^{\tau}), 
\label{SJJ}
\end{equation}
where the Josephson energy is obtained as 
\begin{equation}
E_J\approx 3.7 \frac{t_0^2}{J}. 
\label{EJ}
\end{equation}

\subsection{Resonant Cooper pair tunneling}
\label{LowTBroadening}
 
\begin{figure}[htb]
  \centering
  \includegraphics[width=0.5\textwidth]{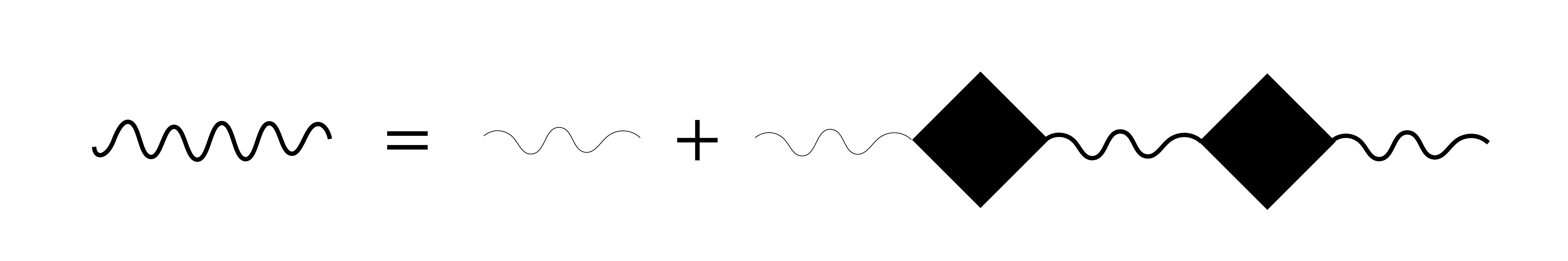}
  \caption{Diagrammatic Dyson equation for Cooperon. Thin wavy lines denote the naked correlation functions of the phase exponents $\mathcal{D}_0(\omega)$, thick wavy lines denote the dressed correlations functions $\mathcal{D}(\omega)$.}
  \label{fig:DysonD}
\end{figure}

In this section we calculate tunneling contribution to the broadening of the Cooper pair resonant state (the Cooperon)  at the energy $\epsilon_1$. The broadening of the Cooper pair resonance is calculated using the Dyson equation for the propagator of the phase correlations, which is represented by diagrams in Fig. \ref{fig:DysonD}.  

\noindent 
The boxes in Fig. \ref{fig:DysonD} correspond to the Josephson couplings given by Eq. (\ref{EJ}). The thin wavy lines denote the correlation function  of phase exponents in the isolated grain without the broadening, which are obtained by the analytic continuation of Eq. (\ref{D_iomega}) to real frequencies $i\omega_n\rightarrow \omega+io$ 
\begin{equation}
\mathcal{D}^R_0(\omega) =  -\frac{2E_1}{\omega^2-\epsilon_1^2+io}, 
\label{DR0}
\end{equation}
where $\epsilon_1=\sqrt{E_1(E_1-2g)}$. 
The thick wavy lines in Fig.\ref{fig:DysonD} denote the correlator of the phase exponents with the self-consistently determined broadening due to the escape of the Cooper pair from the grain. The analytic expression corresponding to Fig. \ref{fig:DysonD} reads 
\begin{equation}
\mathcal{D}^R(\omega)= \mathcal{D}^R_0(\omega)+Z E_J^2\mathcal{D}^R_0(\omega)\left( \mathcal{D}^R(\omega)\right)^2, 
\label{DysonEq}
\end{equation}
where $Z$ denotes the coordination number of the array. 

The formal solution of Eq. (\ref{DysonEq}) can be written as 
\begin{equation}
\mathcal{D}^R(\omega)=\frac{1-\sqrt{1-4ZE_J^2\left(\mathcal{D}^R_0(\omega)\right)^2}}{2ZE_J^2\mathcal{D}^R_0(\omega)}. 
\label{DR_formal}
\end{equation}
Substituting the explicit form Eq. (\ref{DR0}) in Eq. (\ref{DR_formal}), we obtain 
\begin{equation}
\mathcal{D}^R(\omega)=-\frac{4E_1}{\omega^2-\epsilon_1^2+i\sqrt{16ZE_J^2E_1^2-(\omega^2-\epsilon_1^2)^2}}
\label{DR_omega}
\end{equation}
for $|\omega^2-\epsilon_1^2|<4\sqrt{Z}E_JE_1$. Therefore, due to the escape of  Cooper pairs, the resonant  state at the energy $\epsilon_1$ broadens into the semi-circular band. For $|\omega-\epsilon_1|\ll \epsilon_1$ we approximate 
\begin{equation}
\mathcal{D}^R(\omega)\approx -\frac{2E_1/\epsilon_1}{\omega-\epsilon_1+i\left(2\sqrt{Z}E_JE_1/\epsilon_1\right)}
\label{DR_omegaPole}
\end{equation}
For the calculation of conductivity represented in the next Section, we use the simplified expression  
\begin{equation}
\mathcal{D}^R(\omega)=-\frac{4E_1}{(\omega+i\gamma)^2-\epsilon_1^2}.
\label{DR_approx}
\end{equation}
Comparison of Eqs. (\ref{DR_omegaPole}) and (\ref{DR_approx}) for  $\gamma \ll \epsilon_1$ and 
$\omega\approx \epsilon_1$ allows identification 
\begin{equation}
\gamma\approx \frac{2\sqrt{Z} E_J E_1}{\epsilon_1}. 
\label{gamma}
\end{equation}
The solution Eq. (\ref{gamma}) is valid for $\epsilon_1\gg Z^{1/4} \sqrt{E_J E_1}$. This condition is realized at small $E_J$ in the insulating regime with thermally activated conductivity  (see the main text of the paper). 

To evaluate the broadening $\gamma$ close to the critical point for $\epsilon_1\ll Z^{1/4} \sqrt{E_J E_1}$, we consider the equality between  Eqs. (\ref{DR_omega}) and (\ref{DR_approx}) at $\epsilon_1=0$, which results in 
\begin{equation}
\omega^2+i\sqrt{16ZE_J^2E_1^2-\omega^4}=2(\omega^2+2i\omega \gamma-\gamma^2).
\label{gamma-omegaQC}
\end{equation}
To estimate a typical value of $\gamma$ at low frequencies $\omega<2Z^{1/4}\sqrt{E_1E_J}$, we choose  $\omega$ such, that the solution for $\gamma$ is real. Then we obtain from Eq. (\ref{gamma-omegaQC})  
\begin{equation}
\gamma=Z^{1/4}\sqrt{2E_1E_J/3}  
\label{gamma_QC}
\end{equation} 
at $\omega=\sqrt{2}\gamma$. At different values of $\omega$, $\gamma$ acquires an imaginary part, which corresponds to the real part of the self-energy. However, due to the continuity of $\gamma$ as a function of frequency, we accept Eq. (\ref{gamma_QC}) as an estimation of the level broadening at all frequencies for low values of $\epsilon_1$, in particular, in the quantum critical regime.

\subsection{Pair conductivity in the Kuramoto-Josephson array model} 
Here we calculate the conductance between the two Kuramoto grains coupled by the Josephson coupling $E_J$. Assuming that the quantum coherence is lost after a single act of the inter-grain tunneling, the conductivity of the array is obtained form the conductivity of the single junction using the electric circuit theory.  
The action for two Kuramoto grains  reads 
\begin{eqnarray}
\nonumber &&
 S[A, \phi]=\int_0^{\beta}d\tau \sum_{r=1,2}\left[E_1 \sum_{i=1}^N \dot{\phi}_{ri}^2-\frac{g}{N}\sum_{i<j} \cos(\phi_{ri}-\phi_{rj})\right]-\frac{E_J}{N}\sum_{i,j}^N\int_0^{\beta}d\tau \cos\left[\phi_{1i}-\phi_{2j} -2eA(\tau)\right].
 \end{eqnarray}
Here the source vector potential $A(\tau)$ is introduced in such a way that is creates the voltage difference between the two grains ($A=v \tau$ corresponds to a constant voltage $v$ between the grains). The partition function is given by 
\begin{equation}
Z[A]=\int D[\phi_{ir}^{\tau}] e^{-S[A, \phi]}. 
\label{Z[A]}
\end{equation}
The conductivity is calculated as the response to the source vector potential  
\begin{equation}
\sigma=-\lim_{\omega\rightarrow 0} \mathrm{Im}Q(\omega)/\omega.
\label{sigmaQ}
\end{equation}
Here $Q(\omega)$ is the retarded response function, which is obtained by analytic continuation to real frequencies of the  derivative with respect to the source vector potential 
\begin{equation}
Q(i\omega_n)=\frac{\delta^2 \ln Z[A]}{\delta A(i\omega_n)\delta A(-i\omega_n)}\bigg|_{A=0} = \int d  \tau \frac{\delta^2 \ln Z[A]}{\delta A(\tau)\delta A(0)}\bigg|_{A=0} e^{-i\omega_n \tau} \bigg|_{A=0} 
\end{equation}
The derivatives with respect to the source vector potential generate correlation functions of phase exponents, which are given by the following expressions 
\begin{eqnarray}
&& \left\langle \sum_j e^{\pm i \phi_{rj}(\tau)}\right\rangle=0, \\ 
&& \left\langle \sum_{j,j'} e^{ i \phi_{rj}(\tau)} e^{- i \phi_{r'j'}(0)}\right\rangle=N \mathcal{D}(\tau)\delta_{r,r'}. 
\end{eqnarray}
where the Fourier-transform of $\mathcal{D}(\tau)$ is given by Eq. (\ref{D_iomega}). After taking the  derivatives and performing the Fourier transform to Matsubara frequencies, we obtain  the analytic expression for the response function $Q(i\omega_n)$ in the form 
\begin{equation}
Q(i\omega_n)=(2e)^2E_J^2 T\sum_{\Omega_n} \mathcal{D}(i\Omega_n+i\omega_n)\mathcal{D}(i\Omega_n)=
\frac{(2e)^2 E_J^2}{2\pi i} \int_{-\infty}^{\infty} \frac{dx}{e^{\frac{x}{T}}-1}  \left[\mathcal{D}^R(x) -\mathcal{D}^A(x)\right]\left[ 
 \mathcal{D}^R(x+i\omega_n)+ \mathcal{D}^A(x-i\omega_n)\right], 
 \label{Q_iomega}
\end{equation}
where, according to Eq. (\ref{DR_approx})  
\begin{equation}
\mathcal{D}^R(x)=\frac{-4 E_1}{(x-\epsilon_1+i\gamma)(x+\epsilon_1+i\gamma)}.
\label{DRx}
\end{equation}
\begin{figure}[htb]
\vskip -0.5cm
  \centering
  \includegraphics[width=0.3\textwidth]{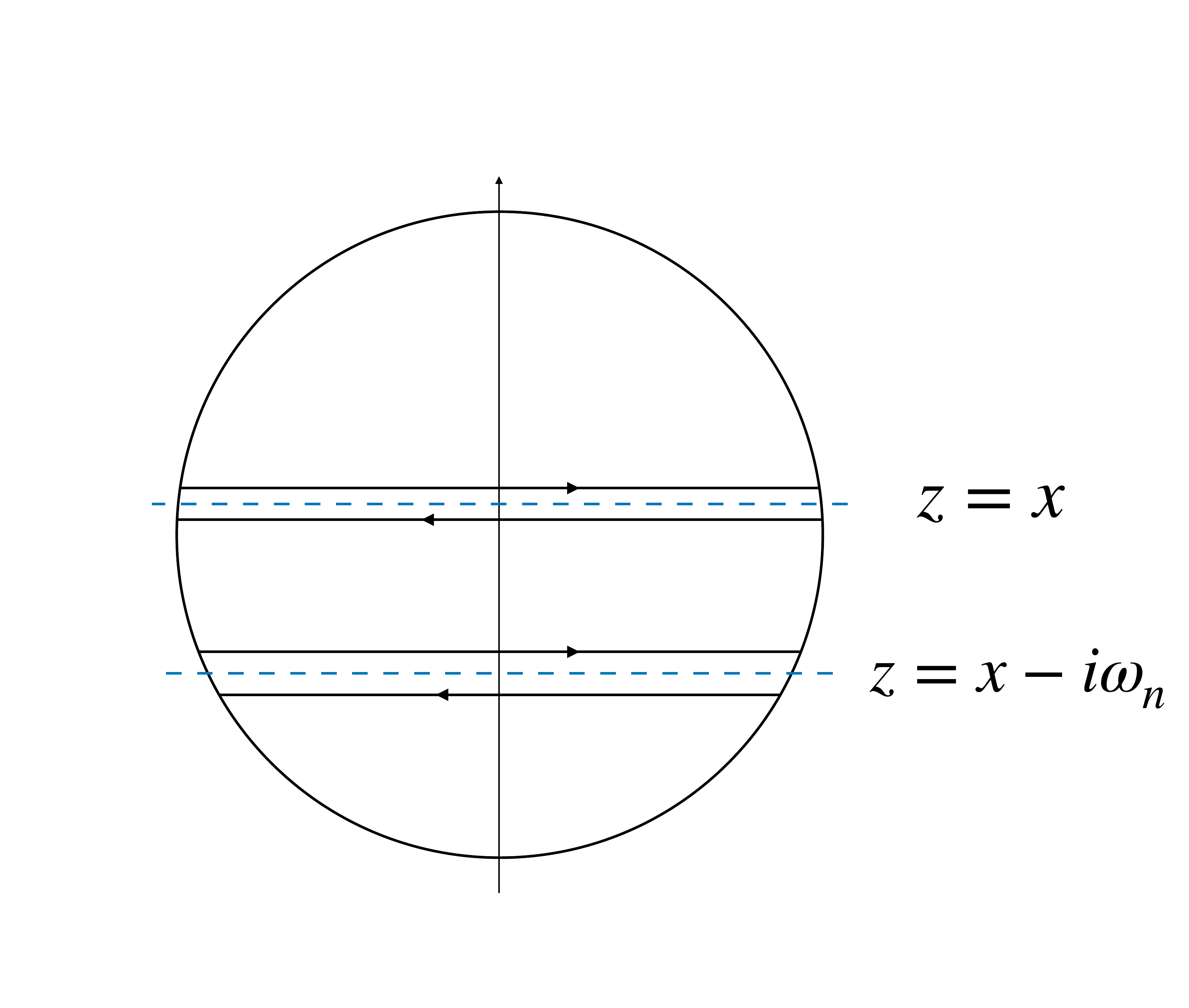}
  \caption{Integration contour in the complex plane $i\Omega_n\rightarrow z$ for calculation of the AL response $Q(i\omega_n)$ in Eq. (\ref{Q_iomega}). Dashed lines denote the branch cuts.}
  \label{fig:IntContourB}
\end{figure}

The response function $Q(i\omega_n)$ is calculated by the analytic continuation of the Matsubara frequency $i\Omega_n$ to the complex plain as shown in Fig. \ref{fig:IntContourB}. Expanding up to the linear term in $i\omega_n$, we obtain 
\begin{equation}
Q(i\omega_n)\approx (2e)^2 E_J^2\frac{ i\omega_n}{2\pi i} \int_{-\infty}^{\infty} \frac{dx}{e^{\frac{x}{T}}-1}   \left(\mathcal{D}^R(x) -\mathcal{D}^A(x)\right) \partial_x\left(\mathcal{D}^R(x)- \mathcal{D}^A(x)\right)=
(2e)^2 E_J^2 
\frac{ \omega_n}{2\pi } \int_{-\infty}^{\infty}
\frac{dx}{8T \sinh^2\left(\frac{x}{2T}\right) } \left(\mathcal{D}^R(x) -\mathcal{D}^A(x)\right)^2, 
\label{Q_omega}
\end{equation}
Finally, performing the analytic continuation to real frequency $i\omega_n\rightarrow\omega+io$ and using Eq. (\ref{sigmaQ}), we obtain the expression for the conductivity (here we restore $\hbar=h/(2\pi)$) 
\begin{equation}
\sigma\! =-(2e)^2
\frac{E_J^2}{h} \int_{-\infty}^{\infty}
\frac{dx}{8T \sinh^2\left(\frac{x}{2T}\right) } \left(\mathcal{D}^R(x) -\mathcal{D}^A(x)\right)^2 =
\! \frac{(2e)^2}{h} \frac{E_J^2 \gamma^2 E_1^2}{T^6} \!
\int_{-\infty}^{\infty}\frac{dy}{\sinh^2 y}\frac{y^2}{\left[\left(y^2-\frac{\epsilon_1^2}{4T^2}-\frac{\gamma^2}{4 T^2}\right)^2 +\frac{\gamma^2 y^2}{T^2}\right]^2}.  
\label{sigmaAL_result}
\end{equation} 
\begin{figure}[htb]
  \centering
  \vskip -0.2 cm
  \includegraphics[width=0.5\textwidth]{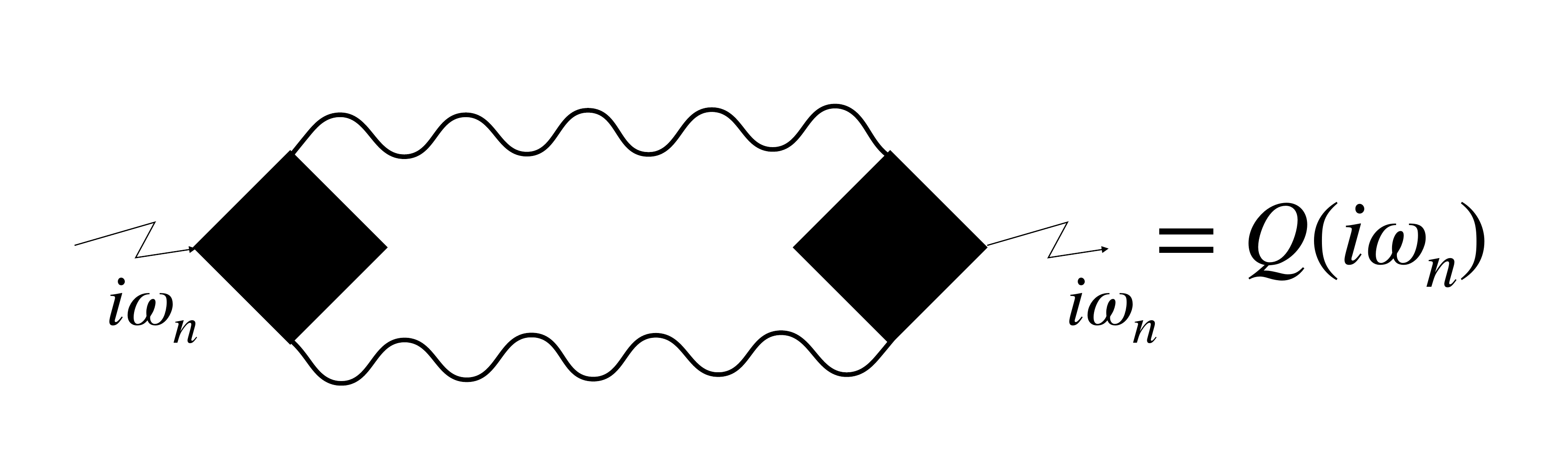}
  \vskip -0.2 cm
  \caption{Aslamasov-Larkin (AL) contribution to the conductivity.}
  \label{fig:Amplitude_supplement}
\end{figure}
 In the insulating phase at low temperatures,  $\gamma\ll \epsilon_1$,  Eq. (\ref{sigmaAL_result}) can be approximated by taking the most singular part of the residue of the second order poles, which results in Eq. (11) in the main text of the paper (here we restore $\hbar=h/(2\pi)$) 
\begin{equation}
\label{sigma_ALmaintext}  
\sigma_{\mathrm{AL}} \! \approx \! \frac{(2e)^2}{h} \frac{8\pi E_1^2 E_J^2}{\gamma\left(\epsilon_1^2  \!+ \!\gamma^2  \right)T  } e^{-\epsilon_1/T} \approx \frac{(2e)^2}{h} \frac{8\pi E_1^2 E_J^2}{\epsilon_1^2 \gamma T  } e^{-\epsilon_1/T}. 
\end{equation}
Close to the quantum phase transition,   for $\epsilon_1\ll \gamma$  the dissipation-induced broadening  cuts off the divergence in Eq. (\ref{sigma_ALmaintext}) for $\epsilon_1\rightarrow 0$.  
Evaluation of the integral in Eq. (\ref{sigmaAL_result}) in the quantum critical regime $\epsilon_1/T\ll 1$ gives 
\begin{eqnarray}
\nonumber && 
\int_{-\infty}^{\infty}\frac{dy}{\sinh^2 y}\frac{y^2}{\left[\left(y^2-\frac{\epsilon_1^2}{4T^2}-\frac{\gamma^2}{4 T^2}\right)^2 +\frac{\gamma^2 y^2}{T^2}\right]^2}\approx 
2\int_{0}^{1}\frac{dy}{\left[\left(y^2-\frac{\epsilon_1^2}{4T^2}-\frac{\gamma^2}{4 T^2}\right)^2 +\frac{\gamma ^2 y^2}{T^2}\right]^2}\approx \left\{ \begin{array}{c}
2(2T/\gamma)^8, \, \, \, \mbox{for} \, \, \, T/\gamma \ll 1, 
\\ 
\frac{5\pi}{16} (2 T/\gamma )^7, \, \, \, \mbox{for} \, \, \, T/\gamma\gg 1,
\end{array} \right.
\end{eqnarray}
which results in the approximate temperature dependence of the conductivity provided in the main text of the paper 
\begin{equation}
\sigma_{\mathrm{AL}}\propto \frac{e^2}{h} \frac{E_J^2 E_1^2 T}{\gamma^6} \left\{ 
\begin{array}{c}
T \, \, \, \mbox{for} \, \,   T/\gamma \ll 1, \\ 
\gamma  \, \, \, \mbox{for} \, \,  T/\gamma \gg 1. 
\end{array}\right.
\label{sigmaAL_QCrit_sup}
\end{equation}

\subsection{Escape rate of a Cooper pair in the incoherent JJ-array}
The array of Josephson junctions is determined by the conventional Hamiltonian 
\begin{equation}
H_{\mathrm{JJ}}=\frac{E_C}{2}\sum_r \partial^2_{\phi_r}-E_J \sum_{r,r'} \cos(\phi_r-\phi_{r'}),
\label{HJJ} 
\end{equation}
where the sum runs over the nearest neighbor grains. 
The regime of the incoherent Josephson array is characterized by the relation $T\gg E_J\gg E_C$, where $T$ denotes the temperature, $E_J$ is the Josephson energy, and $E_C$ is the charging energy of a single superconducting grain. In that regime, each grain has a well defined local superconducting order, although the phases of different grains are uncorrelated due to large temperature fluctuations. It is then reasonable to adopt a set of isolated grains at finite temperature as a zero approximation and treat the Josephson coupling as a perturbation. The eigenstates of the isolated grain are quantized according to the angular momentum $\ell$ canonically conjugated to the superconducting phase $\phi$ (physically those states correspond to a fixed number of Cooper pairs in the grain). At high temperature, each grain is in the mixed state that is characterized by the diagonal density matrix determined by the thermodynamic Boltsmann distribution    
\begin{equation}
P_{\ell}=\exp[-E_C\ell^2/(2T)]/\mathcal{N}, 
\label{P_ell}
\end{equation}
where the normalization 
\begin{equation}
\mathcal{N}=\sum_{\ell=-\infty}^{\infty} e^{-E_C\ell^2/(2T)} \approx \sqrt{\frac{2\pi T}{E_C}}. 
\label{Zresult}
\end{equation} 
The Josephson coupling acts as a hopping amplitude for Cooper pairs between neighbor grains. Because the superconducting phases of different grains are uncorrelated, the escape of a Cooper pair from the grain results in the relaxation of the superconducting phase.  

In this section we estimate the escape rate of a Cooper pair from a grain as the broadening of the angular momentum energy levels. The latter can be evaluated  self-consistently, using the Dyson equation (see Fig. \ref{fig:DysonD}). For the retarded Green function of the angular momentum $\ell$, the Dyson equation reads 
\begin{equation}
\mathcal{D}^R_{\ell}(\omega)=D^R_{\ell}(\omega)+D^R_{\ell}(\omega)E_J^2 Z \left(\sum_{\ell'} P_{\ell'} \mathcal{D}^R_{\ell'}(\omega)\right) 
\mathcal{D}^R_{\ell}(\omega) 
\label{DysonDl}
\end{equation}
Here $D^R_{\ell}(\omega)=\left(\omega-\delta_{\ell}+io\right)^{-1}$ denotes the retarded Green function of the single superconducting grain in the excited state with the angular momentum $\ell$, $\delta_{\ell}=E_C(\ell+1/2)$, and $Z$ denotes the coordination number. The function $D^R_{\ell}(\omega)$ is easily obtained from Eq. (\ref{Dl_0}) by the replacement $E_1\rightarrow E_C/2$.   $P_{l'}$ denotes the probability of a grain to be thermally activated in the state with the angular momentum $l'$ as given by Eq. (\ref{P_ell}).  Assuming for the exact Green function in the form 
\begin{equation}
\mathcal{D}^R_{\ell}(\omega)=\frac{1}{\omega-\delta_{\ell}+i\gamma}
\label{Dlexact}
\end{equation}
one obtains from Eq. (\ref{DysonDl}) the self-consistency condition for $\gamma$, which reads  
\begin{equation}
1=E_J^2 Z \sum_{\ell} \frac{P_{\ell}}{\delta_{\ell}^2+\gamma^2}.
\label{gamma_sc-highT}
\end{equation}
Substituting explicit expressions $P_{\ell}=\sqrt{\frac{E_C}{2\pi T}}\exp[-
E_C\ell^2/(2T)]$, $\delta_{\ell}=E_C(\ell+1/2)$, we estimate the sum in Eq. (\ref{gamma_sc-highT}) in the  high-temperature regime $E_C/T\ll 1$ replacing it by the integral, thus obtaining 
\begin{equation}
1\approx  \frac{Z E_J^2}{\sqrt{2\pi E_C T}} \int dx \frac{\exp[-x^2/(2E_C T)] }{x^2+\gamma^2}  =\frac{\sqrt{\pi} Z E_J^2}{\sqrt{2 E_C T}\gamma} \exp\left(\frac{\gamma^2}{2E_CT}\right)\mathrm{erfc} \left(\frac{\gamma}{\sqrt{2 E_C T}}\right),  
\label{gamma_selfcons}
\end{equation}
where $\mathrm{erfc}(x)$ denotes the complementary error function. 
Finally, we represent the self-consistent equation for the level broadening in the form 
\begin{equation}
\gamma=\frac{\sqrt{\pi} Z E_J^2}{\sqrt{2 E_C T}}\exp\left(\frac{\gamma^2}{2E_CT}\right) \mathrm{erfc} \left(\frac{\gamma}{\sqrt{2 E_C T}}\right). 
\label{gamma_highT}
\end{equation} 
Eq. (\ref{gamma_highT}) can be solved explicitly in the two limit cases using the asymptotic behavior 
\begin{equation}
\exp \left(x^2\right) \mathrm{erfc} (x) \approx \left\{
\begin{array}{c}
\frac{1}{\sqrt{\pi}x}, \, \, \, \mbox{for}, x\gg 1, \\ 
1, \, \, \, \mbox{for}, x\ll 1.
\end{array}\right. 
\end{equation}
We obtain 
\begin{eqnarray}
&& 
\gamma\approx \frac{\sqrt{\pi}Z E_J^2}{\sqrt{2E_C T}}, \, \, \mbox{for} \, \, E_J^2\ll E_C T, 
\label{gamma_EJsmall}\\ 
&& 
\gamma\approx \sqrt{Z} E_J, \, \, \mbox{for} \, \, E_J^2\gg E_C T.  
\label{gamma_EJlarge}
\end{eqnarray}

\subsection{Conductivity in the incoherent Josephson array}
In this section we evaluate the temperature dependence of conductivity for the  incoherent Josephson array at temperatures much exceeding the superconducting transition temperature $T\gg  E_J$. We adopt the picture of the array as an electric circuit of Josephson junctions, for which the total conductivity can be calculated from the conductivity of a single junction using circuit theory  rules.   Therefore, the  temperature dependence of the total conductivity is determined by that of the single Josephson junction. The dynamics of the relative superconducting phase of the Josephson junction is governed  by the action 
\begin{equation}
S_{JJ}=\int dt \left\{\frac{C\dot{\phi}^2}{2(2e)^2} +  E_J \cos(\phi)\right\}, 
\end{equation}
where $C$ denotes the capacitance of the junction related to the charging energy in Eq. (\ref{HJJ}) by $E_C=(2e^2)/C$. For a Josephson junction embedded in the incoherent Josephson array, the equation of motion for the superconducting phase should be supplemented by terms describing the dissipation and thermal noise, so that the phase dynamics is governed by the Langevin equation 
\begin{equation}
C\ddot{\phi} + \sigma \dot{\phi}=(2e)^2  E_J \sin(\phi)+\xi(t). 
\label{LangevinPhase_sup}
\end{equation}
Here $\sigma$ introduces the phase relaxation and $\xi(t)$ is the Langevin source related to the relaxation by the fluctuation-dissipation relation 
\begin{equation}
\langle \xi(t) \xi(t')\rangle=2 \sigma T \delta(t-t'). 
\label{FDT_sup}
\end{equation}
Taking into account the relations between the phase $\phi$, the voltage $v$, and the total charge of the grain $Q$,  $\dot{\phi}=2e v$,  $Q=Cv$, Eq. (\ref{LangevinPhase_sup}) can be rewritten in form of the 
 prominent resistively and capacitively shunted junction (RCSJ) 
 \begin{equation}
 I_C+I_R+I_S=\xi(t)/(2e), 
 \label{RCSI0}
 \end{equation}
 where $I_C=C\dot{v}=\frac{C}{2e}\ddot{\phi}$, $I_R=v/R=\frac{\sigma}{2e}\dot{\phi}$, and $I_S=-2e  E_J \sin(\phi)$. Here we assume that the phase coherence of a Cooper pair is lost along any indirect path  connecting the two grains of the junction. Therefore, the escape of a Cooper pair into the rest of the array acts as the source of the resistive current $I_R$. 
For the current-biased RCSJ, rhs of Eq. (\ref{RCSI0}) should be amended with the external current 
  \begin{equation}
 I_C+I_R+I_S=\xi(t)/(2e)+I, 
 \label{RCSI}
 \end{equation}
 In turn, the  external dc  current can be eliminated from RCSI equation Eq. (\ref{RCSI}) by stepping back to the representation in terms of phase variables and performing the gauge transformation $\phi\rightarrow\phi+2ev t$, where $v=I/\sigma$. This transformation reveals $\sigma$ as the conductivity of the RCSI junction. After the gauge transformation the equation of the current biased RCSI junction becomes 
 \begin{equation}
 I_C+I_R+\tilde{I}_S=\xi(t)/(2e), 
 \label{RCSI_gauged}
 \end{equation}
where  $\tilde{I}_S=-2e  E_J \sin(\phi+2ev t)$. The relation of the parameter $\sigma$  to the escape rate of a Cooper pair out of the junction into the rest of the array can be clarified by considering the time derivative of Eq. (\ref{RCSI_gauged}).  Using the relations $\dot{I}_R=\frac{\sigma}{2e}\ddot{\phi}=\frac{\sigma}{C}I_C$, one represents the equation for the time-derivative of the current in the form 
\begin{equation}
\dot{I}_C+\dot{\tilde{I}}_S=-\frac{\sigma}{C} I_C+ \frac{1}{2e} \dot{\xi}(t). 
\label{LangevinCurrent}
\end{equation}
After averaging  over the thermal fluctuations for temperatures much larger than the superconducting transition temperature in the array, $T\gg  E_J$, one can neglect the contribution of the superconducting current, and obtain the equation for the dynamics of the $RC$ junction
\begin{equation}
\dot{I}_C=-\frac{\sigma}{C} I_C. 
\label{RC}
\end{equation}
 Therefore, the parameter  $C/\sigma$ constitutes the time constant of the $RC$ junction, which in our case should be associated with the escape rate $\gamma$ of a Cooper pair out of the grain into the rest of the array, $\gamma=\sigma/C$. 
The parameter  $\gamma$ is determined by  Eq. (\ref{gamma_highT}).  Finally, using the relation between the capacitance and the charging energy, $E_C=(2e)^2/C$, we obtain 
\begin{eqnarray}
&& 
\sigma\sim \frac{Z E_J^2}{E_C^{3/2}\sqrt{ T}}, \, \, \mbox{for} \, \, E_J^2\ll E_C T, 
\label{sigma_EJsmall}\\ 
&& 
\sigma\sim \sqrt{Z} E_J/E_C, \, \, \mbox{for} \, \, E_J^2\gg E_C T.  
\label{sigma_EJlarge}
\end{eqnarray}

\end{document}